# Approcher l'identité professionnelle d'enseignants universitaires de physique : un levier pour initier des changements de pratiques pédagogiques

**Cécile de Hosson**
Université Paris Diderot, laboratoire de didactique André Revuz

**Nicolas Décamp**
Université Paris Diderot, laboratoire de didactique André Revuz

**Émilie Morand**
STRAVIA, université Paris Descartes, CERLIS

**Aline Robert**
Université Cergy-Pontoise, laboratoire de didactique André Revuz

RÉSUMÉ • À la suite de profondes modifications dans les programmes de physique-chimie de lycée au cours de ces trois dernières années, l'UFR de physique de l'université Paris-Diderot a souhaité élaborer un projet de rénovation de ses modalités d'enseignements. À la faveur de ce projet, nous avons pu questionner 104 enseignants-chercheurs de cette UFR par l'intermédiaire de 23 entretiens exploratoires et de 81 questionnaires. Nous les avons sondés sur les changements possibles et souhaitables des modalités pédagogiques et d'organisation (en tenant compte des difficultés actuelles rencontrées par les étudiants). Cette étude nous a permis de dégager des éléments de leur identité professionnelle en tant qu'enseignants, identité que nous avons explorée selon différents axes : règles du métier, qualités et compétences nécessaires à son exercice, valeurs du métier, ainsi qu'un axe didactique spécifique à l'enseignement de la physique. L'analyse des données recueillies nous permet de dégager de grandes tendances parmi les conceptions de l'enseignement des enseignants-chercheurs interrogés, comme le fait qu'un bon enseignant est avant tout un bon physicien, ou que l'enseignement doit favoriser l'interaction avec les étudiants. Elle nous permet également de mettre en lumière des tensions entre ce que les enseignants chercheurs interrogés souhaiteraient faire dans leur enseignement et ce qu'ils déclarent faire en pratique ou pensent réalisable.

ABSTRACT • Approaching physics teachers' professional identity in higher education: a lever for initiating changes in teaching practices

As a result of the deep modifications of the French physics and chemistry curricula in upper secondary school during these last three years, the physics department of the Université Paris Diderot (France) has developed a renovation project concerning its methods of teachings. As science education researchers we took part in this project questioning 104 university academics of this department through 23 exploratory interviews and 81 questionnaires. They have been asked about the possible and desirable changes concerning pedagogical methods and organization (with respect to current difficulties met by the students). This study has enabled us to pinpoint elements of their professional identity as teachers. This identity has been explored according to various directions:rules which govern their profession, qualities and skills for the practice, values of the profession, as well as a didactic dimension, dealing specifically with physics' teaching. The analysis of the collected data allows us to identify major trends among university academics' conceptions about teaching, such as: a good teacher is above all a good physicist, or, teaching has to favor interactions with students. It also allows us to bring to light tensions between what university academics would like to do in their teaching and what they declare to do in practice or think feasible.

# Introduction

Les programmes de physique-chimie au lycée, en France (classes de seconde, de première S et de terminale S) ont subi ces trois dernières années de profondes modifications. Celles-ci touchent à la fois aux intentions, aux contenus ainsi qu'aux pédagogies encouragées par l'Institution[1] : les savoirs de la physique sont abordés à l'occasion d'explorations thématiques engageant des domaines transversaux tels que le sport, la santé, etc. ; leur formulation se veut essentiellement qualitative, le formalisme mathématique, support d'expression des lois et des principes en physique, apparaissant fortement réduit ; les activités expérimentales sont, dans la mesure du possible, abordées sous la forme de « démarches d'investigation »[2] et les activités documentaires s'installent de manière prépondérante[3]. Il s'agit, en résumé, de placer l'élève en situation d'acquérir les bases d'une « culture scientifique ».

---

1   *Bulletin officiel de l'Éducation nationale* [dorénavant BOEN], spécial n°8 du 13 octobre 2011 (classe de terminale S), *BOEN*, spécial n°9, 30 septembre 2010 (classe de première S), *BOEN*, spécial n°4, 29 avril 2010 (classe de seconde).

Compte tenu de ces nouvelles orientations curriculaires, les bacheliers qui ont intégré les cursus universitaires scientifiques (ceux de physique, en particulier) depuis la rentrée 2013 ont un profil assez différent de celui qui caractérisait les bacheliers formés selon les programmes en place depuis 2002[4]. Cette situation, concomitante au constat de faibles taux de réussite des étudiants de première année[5] a conduit l'UFR de physique de l'université Paris-Diderot (France) à s'interroger sur les conditions d'élaboration d'un projet de rénovation des modalités d'enseignement.

En tant que chercheurs en didactique de la physique, enseignants de l'UFR de physique de l'université Paris-Diderot, nous avons souhaité accompagner l'élaboration de ce projet en explorant, avec l'aide d'une chercheuse en sociologie du travail et d'une didacticienne des mathématiques, l'espace des changements possibles et souhaitables en terme de modalités pédagogiques (encadrement, évaluation, etc.) et dans les dispositifs organisationnels (CM, TD, etc.), tenant compte des difficultés actuelles rencontrées par les étudiants. Pour cela nous avons réalisé une enquête auprès des enseignants-chercheurs (désignés ci-après EC) de cette UFR dans le but de définir les termes d'un projet de rénovation acceptable, consensuel et fédérateur. Au-delà de sa dimension prospective (visant à définir les changements envisageables en vue d'une meilleure réussite des étudiants), cette enquête présente un caractère heuristique puisqu'elle est l'occasion de remonter à des éléments d'identité professionnelle d'enseignants-chercheurs considérés dans leur métier d'enseignant. Nos connaissances didactiques nous permettent, dans ce cadre, d'introduire dans certains de nos questionnements et dans nos interprétations, des éléments directement liés aux contenus de savoir en jeu (ici, la physique).

---

2   Dans le prolongement de l'école primaire et du collège.

3   Les programmes de sciences physiques de terminale accordent une place « centrale » à deux compétences : « extraire et exploiter des informations ». Celles-ci se voient déclinées au long du programme sous la forme d'« études documentaires » et d'« analyses et synthèses de documents » (*BOEN*, spécial n°8 du 13 octobre 2011). Ces compétences ont été largement valorisées dans le sujet de sciences physiques de l'épreuve du baccalauréat S de 2013.

4   Le document : <http://www2.ac-lyon.fr/enseigne/physique/phychi2/IMG/pdf/Profils_en_physique-chimie_des_bacheliers_scientifiques_a_partir_de_2013.pdf>, réalisé par l'inspection générale de physique présente ces nouveaux profils « bachelier S » en physique et en chimie.

5   En 2012 la moyenne à l'examen de physique du premier semestre de l'ensemble des 331 étudiants inscrits en L1 de sciences exactes à l'université Paris-Diderot était de 8,04/20. 75 % de ces étudiants avaient une note inférieure à 10/20.

## 1. Définition d'éléments d'identité professionnelle d'enseignants-chercheurs de physique : problématique et cadre de recherche

### 1.1. L'enseignement supérieur : un objet de recherche

Les recherches sur les pratiques pédagogiques et l'identité professionnelle des enseignants dans l'enseignement supérieur prennent une place de plus en plus importante dans le paysage des recherches en éducation. Dans un ouvrage collectif paru il y a près d'une dizaine d'années, Annoot et Fave-Bonnet présentaient la première synthèse des questions associées à l'étude des pratiques pédagogiques à l'université (Annoot & Fave-Bonnet, 2004). Il s'agissait pour les auteurs d'apporter un éclairage inédit sur les pratiques quotidiennes des enseignants et des étudiants, sur leurs représentations et sur la portée de leurs actions. Les contributions qui structuraient cette synthèse ouvraient la voie au développement de travaux à initier et/ou à poursuivre ; l'exploration des pratiques des enseignants-chercheurs formait l'une de ces pistes et entendait interroger l'idée de culture commune à travers l'étude de questions associées par exemple au traitement de l'hétérogénéité des étudiants (Altet, 2004), aux écarts entre le déroulement prévu et le déroulement effectif du cours (Trinquier & Terrisse, 2004), ou encore aux éléments (notamment institutionnels) qui contraignent les pratiques (Clanet, 2004), éléments qui depuis lors, ont continué à être questionnés (Langevin, 2008 ; Rege-Collet & Berthiaume, 2009).

Plus précisément, l'identité professionnelle des enseignants-chercheurs est interrogée par les recherches tournées vers les pratiques pédagogiques à l'université. L'enseignant-chercheur est chercheur avant d'être enseignant ; il a été formé comme chercheur et demeure évalué comme tel (Fave-Bonnet, 1999 ; Norton *et al.*, 2005 ; Musselin, 2008 ; Becerra Labra *et al.*, 2012). Comme le soulignait Endrizzi dans un dossier de l'Institut français de l'Éducation consacré à la formation pédagogique des enseignants du supérieur : « les enseignants-chercheurs sont physiciens avant d'être enseignants de physique » ; leur identité professionnelle est fondée sur « l'allégeance » à leur discipline de recherche (Endrizzi, 2011). Ce constat a ouvert la voie à de nombreuses initiatives locales de développement professionnel (c'est-à-dire : actions de formation continue) des enseignants-chercheurs dont certaines font l'objet d'évaluation[6].

Les dispositifs organisationnels de l'enseignement supérieur font également l'objet de plusieurs études. Ainsi, à propos des cours magistraux (CM), Clanet (2004) indique que les enseignants-chercheurs disent maintenir et rechercher l'activité des étudiants par le jeu de questions-réponses, d'humour, d'appels à des situations de la vie courante, etc. Mais ces stratégies ne semblent pas repérées par les étudiants. D'ailleurs, précise Clanet, « dans 70 % des cas, les questions restent sans réponse » (Clanet, 2004, p. 108). Dans cette même enquête, Clanet relève une plus grande variété des pratiques dans les séances de travaux dirigés (TD), séances au sein desquelles l'enseignant cherche à s'enquérir de la compréhension, à favoriser l'interactivité, etc. Dans le contexte particulier de l'enseignement de la physique, on relèvera l'étude exploratoire conduite par Henderson et Dancy (2007) qui questionne les liens entre contenus enseignés et modalités correspondantes. À travers l'analyse d'entretiens individuels de six enseignants de physique nord-américains, Henderson et Dancy ont mis en évidence les obstacles susceptibles de heurter la mise en place de stratégies d'enseignement en rupture avec ce que les chercheurs qualifient d'enseignement traditionnel (cours magistral illustré par des exercices corrigés par les enseignants), alors même que les résultats de la recherche en didactique de la physique (PER-Physics Education Research) pénètrent progressivement la sphère universitaire nord-américaine (Henderson & Dancy, 2007)[7]. Ils sont (re)connus des enseignants interviewés qui expriment se trouver en tension entre ce qu'il conviendrait de faire et ce qu'ils font. On retiendra de cette étude le poids de l'organisation institutionnelle habituelle et répétée sur la persévérance des pratiques ascendantes où enseigner c'est « dire le savoir » (Clanet, 2001).

---

6  Un regard sur les actes du Congrès international de pédagogie universitaire AIPU permet de prendre la mesure de l'intérêt croissant de la communauté universitaire en faveur des pédagogiques innovantes (TICE, MOOC, pédagogie par projet, etc.), de la formation et de l'accompagnement des pratiques pédagogiques, des questions de développement professionnel, etc.

### 1.2. La sociologie du travail : un cadre pour approcher des éléments d'identité professionnelle d'enseignants du supérieur

Dans son orientation heuristique (et en écho aux enquêtes précédentes), notre enquête vise à caractériser des éléments d'identité professionnelle d'enseignants-chercheurs de physique, préalable indispensable selon nous pour identifier les conditions de réussite des changements qui seront envisagés par la suite. Du point de vue de la sociologie du travail, l'identité professionnelle peut se définir comme « un ensemble d'éléments particuliers de représentations professionnelles, spécifiquement activé en fonction de la situation d'interaction et pour répondre à une visée d'identification/différenciation avec des groupes sociétaux ou professionnels » (Blin, 1997). Dans le cas particulier de l'enseignement, Cattonar (2001) définit l'identité professionnelle enseignante comme la façon dont un individu « enseignant » se définit dans son rapport avec sa pratique professionnelle d'enseignant : « ce sont les caractéristiques qui l'identifient en tant qu'enseignant et que l'enseignant partage, qu'il a en commun avec d'autres enseignants du fait d'appartenir au même groupe professionnel » (Cattonar, 2001, p. 5). Nous envisageons ici le groupe professionnel « enseignant-chercheur » comme formant une « subculture » caractérisée par des modes de percevoir, de penser et d'agir particuliers, par des normes, des valeurs et des règles propres qui sont liés à leur objet de travail et à leur pratique professionnelle (Cattonar, 2001, p. 6). L'identité professionnelle enseignante n'est donc pas une identité structurée et stable, extérieure à l'enseignant, qu'il recevrait/adopterait passivement mais une construction où interviennent sa subjectivité, ses propres représentations, motivations et intérêts. En perpétuels mouvements, cette construction évolue en fonction de ses expériences et de la confrontation de ses valeurs avec celles de sa communauté. C'est une sorte de compromis entre les «besoins et les désirs de l'individu et les valeurs des différents groupes avec lesquels il entre en relation» (Dubar, 1996). Il convient de préciser que l'on parle ici d'identité professionnelle de l'enseignant-chercheur dans son travail d'enseignant. D'un point de vue conceptuel, et conformément aux éléments de définition précisés ci-dessus, nous avons retenu plusieurs dimensions de l'identité professionnelle :

– les règles du métier (ce que l'enseignant juge légitime pour «bien» exercer son métier en termes de règles, de comportements, de fonctionnement, et ce qui, au contraire, lui semble illégitime, peu approprié) ;

– les qualités et les compétences nécessaires à l'exercice du métier (les qualités et compétences que l'enseignant mobilise pour exercer son métier) ;

– les valeurs du métier (ce que l'enseignant valorise dans son métier, les fonctions qu'il aimerait déléguer et celles qu'il ne déléguerait jamais, constitutives de son cœur de métier).

---

7 On notera à ce propos que la revue de recherche *Physical Review* consacre une partie de sa politique éditoriale à la recherche en didactique de la physique (physics education research). De même en est-il de la revue *Science*.

Nous ajoutons une dimension «didactique» aux précédentes, qui nous permettra d'éclairer certains aspects du rapport que les EC entretiennent, non pas au savoir en général mais à la physique en particulier. Cela servira à examiner la façon dont les EC envisagent le rapport au savoir des étudiants (par exemple, les raisonnements pouvant faire obstacle à l'apprentissage, etc.). Il s'agit en effet de pouvoir analyser et caractériser des éléments spécifiques de l'identité scientifique des EC considérés en tant que physiciens.

D'un point de vue méthodologique, notre enquête se donne pour objectif de mettre à jour des conceptions d'enseignants-chercheurs de physique relatives à chacune de ces dimensions de l'identité professionnelle. Le terme «conceptions» est ici considéré au sens de Langevin (2008) comme une «attitude mentale qui permet à l'individu d'appréhender et d'interpréter la réalité». Afin d'affiner nos interprétations, nous ajoutons aux dimensions précédentes :

– des éléments généraux de compréhension (nombre d'années d'exercice, cursus, formation, etc.) ;

– des questions sur des changements possibles (touchant à ce que les enseignants ont déjà fait, à ce qu'il serait possible de faire, etc.).

Nous avons pris le parti d'élaborer un questionnement double, composé de questions directement en lien avec la commande qui nous revenait, et des questions visant à identifier de manière plus large, les éléments d'identité professionnelle des EC, les conceptions sur lesquelles s'appuie leur pratique d'enseignement. Ceci nous permettant de définir les conditions d'acceptabilité des changements qui pourraient être envisagés après l'enquête.

## 2. Recueil de données

Nous avons choisi de conduire notre enquête selon une exploration mixte en deux temps engageant deux corpus contribuant chacun aux dimensions prospectives et heuristiques de notre enquête. Le premier est constitué de vingt-trois transcriptions d'entretiens exploratoires approfondis, le second de réponses à 81 questionnaires auto-administrés remplis en ligne, anonymes, construits à partir de l'analyse des entretiens.

### 2.1. Les entretiens

Prévus pour une durée moyenne de 40 minutes, ils étaient organisés de façon à recueillir l'avis des enseignants-chercheurs interviewés sur les aspects suivants, présentés par l'interviewer au début de l'entretien, rappelés au fil de l'entretien et introduits par « que penses-tu de… » :
– L'organisation traditionnelle cours/TD/TP ;
– Les pratiques innovantes ;
– Les difficultés des étudiants ;
– L'évaluation : des étudiants et des enseignements ;
– La formation des enseignants du supérieur.

Les interviewés sont les vingt-trois premiers enseignants-chercheurs à avoir répondu positivement à une sollicitation collective par mail au mois d'octobre 2012[8]. L'échantillon se compose de deux allocataires-moniteurs, treize maîtres de conférences, huit professeurs des universités. Le temps moyen de parole laissé à la charge de l'interviewé est compris entre 70 et 80 % de la durée totale de l'entretien ; peu de relances ont été nécessaires.

### 2.2. Les questionnaires

Le questionnaire (23 questions à choix multiple non exclusif, sans possibilité de priorisation, avec demandes de justification, hors questions relatives au profil des répondants) a été construit à partir de l'analyse des entretiens. Les invariants repérés dans les *verbatim* (principes partagés par les enseignants-chercheurs interrogés) sont devenus des *items* des questions à choix multiples, de même que les propositions d'innovations et de changements. Chacune des questions renvoie à l'une des dimensions de l'identité professionnelle telles que nous les avons définies plus haut (voir tableau 1).

---

8   Envoi d'un mail collectif par le Conseil des enseignements de l'UFR de physique, octobre 2012.

| Dimension questionnée | Définition | Questions |
|---|---|---|
| Règles | Ce que l'enseignant juge légitime pour bien exercer son métier en termes de règles, de comportements, de fonctionnement, et ce qui, au contraire, lui semble illégitime, peu approprié | Quel est l'objectif d'un cours en amphithéâtre ? Comment s'assurer que l'objectif a été atteint ? Quel est l'objectif d'un TD ? Comment s'assurer que l'objectif a été atteint ? Quelle proposition vous semble la mieux adaptée à l'évaluation des étudiants ? Qu'est-ce qu'un contrôle continu efficace ? |
| Qualités | Les qualités et compétences que l'enseignant mobilise pour exercer son métier | Que trouvez-vous difficile dans votre métier d'enseignant ? Ressentez-vous le besoin d'être formé en tant qu'enseignant ? |
| Valeurs | Ce que l'enseignant valorise dans son métier, les fonctions qu'il aimerait déléguer et celles qu'il ne déléguerait jamais, constitutives de son cœur de métier | Qu'est-ce qui vous plaît le plus dans votre métier d'enseignant-chercheur ? Que seriez-vous prêt à déléguer/à ne pas déléguer ? |
| Éléments didactiques | Ce qui distingue l'enseignant « physicien » d'un enseignant d'une autre discipline de la physique et qui renseigne sur son rapport à la physique d'une part, au rapport à la physique des étudiants, d'autre part | Que trouvez-vous difficile dans votre métier d'enseignant ? Quelles sont les sources de difficulté des étudiants ? |
| Éléments généraux | Éléments concernant le parcours de l'enseignant et critères sociodémographiques | Sexe, corps, nombre d'années d'exercice, cursus ? Comment avez-vous appris à enseigner ? |
| Questions visant à tester virtuellement des changements possibles | Propositions de changement possible, de nouvelles modalités, etc. | La modalité cours en amphithéâtre est remplaçable par... La modalité TD est remplaçable par... Seriez-vous favorable à la mise en place de séances de tutorat ? Seriez-vous favorable à une modalité cours-TD exclusive ? À quelle modalité d'évaluation de votre enseignement seriez-vous favorable ? Seriez-vous favorable à des séances d'autoconfrontation ? Seriez-vous favorable à l'utilisation de *clickers* ? |

Tabl. 1 : répartition des questions (du questionnaire) selon les dimensions
de l'identité professionnelle des enseignants-chercheurs interrogés
À noter, certaines questions ne sont pas exclusives d'une dimension.

Ce second corpus est constitué des réponses de 81 enseignants-chercheurs (60 hommes, 21 femmes)[9]. L'échantillon représente environ 40 % de la population sollicitée (210 enseignants-chercheurs permanents et allocataires moniteurs), mais l'on notera une forte hétérogénéité de la répartition de ce taux de participation selon le corps d'appartenance des enseignants-chercheurs ayant répondu au questionnaire : près des trois quarts des professeurs des universités sollicités ont répondu au questionnaire alors que moins d'un tiers des maîtres de conférences sollicités l'ont renseigné (le taux de réponse des moniteurs est quant à lui très faible : un cinquième des allocataires-moniteurs sollicités).

## 3. Méthodologie d'analyse des données

### 3.1. Les entretiens

L'intention des entretiens étant de produire des *items* pour un questionnaire ultérieur, nous avons choisi d'analyser les vingt-trois transcriptions d'entretien en procédant par analyse de contenu (Bardin, 1971). Plus particulièrement, nous avons repéré (pour les catégoriser) les « phrases récurrentes », celles qui circulent et « qui restent inchangées d'un individu à l'autre » et qui « correspondent à des processus sous-jacents essentiels » ; celles qui « fondent le sens commun autour d'une question » (Kaufmann, 2011, p. 96).

Prenons un exemple. Voici la réponse de l'enseignant E2 à la première question « que penses-tu de la modalité d'organisation cours/TD/TP ? ». Nous soulignons en gras les mots ou expressions qui ont servi de support à l'encodage puis à la catégorisation :

> « *Ce qui est commun et négatif à l'ensemble de ces trois modalités, c'est le fait que dans les trois cas, les étudiants sont passifs. Il n'y a **aucune interaction** avec l'enseignant et en gros, un livre suffirait en cours et des exercices corrigés suffiraient en TD. En TP, c'est un peu différent, par exemple en PhysExp parce que les étudiants sont autonomes dans la résolution de problèmes complexes. Mais dans leur forme traditionnelle, les TP laissent les étudiants **passifs** : il faut juste qu'ils suivent un protocole élaboré en dehors d'eux. Moi, je **serais favorable à des cours-TD** dans lesquels le cours serait donné pour ensuite être investi dans des exercices **qui seraient très peu nombreux**. Il faudrait que les enseignants mettent **un poly de leur cours** à disposition des étudiants et s'ils se réfèrent à un livre, il faudrait que leur cours suive le livre parce que sinon, ça demande trop d'efforts à l'étudiant pour relier les deux.* »
> [enseignants-chercheurs E2]

On repère ici les mots et/ou expressions en lien avec l'idée de passivité (code : PAS) qui renvoie à la catégorie « attitude/difficultés des étudiants ». On retient également de cet extrait une proposition alternative d'organisation (le cours-TD) et une recommandation en direction d'un usage plus systématique d'un polycopié de cours (ou d'un ouvrage), deux éléments qui seront proposés dans le questionnaire.

---

[9] Enseignants sollicités par mail collectif entre décembre 2012 et janvier 2013 (relance début février 2013) par le Conseil des enseignements de l'UFR de physique.

Le nombre d'exercices par feuille de TD sera conservé comme une variable à laquelle nous associerons le travail de l'étudiant tel que perçu par les EC interrogés.

### 3.2. Les questionnaires

Les réponses au questionnaire sont analysées à l'aide d'un tableur. Au-delà des données statistiques brutes (qui permettent d'obtenir des tendances, notamment pour les questions associées à des réponses binaires de type « oui » ou « non »), on procède également par tris (croisement de variables) afin de dégager des cohérences et des contradictions (pour un même individu, par exemple, ou pour des individus ayant le même type de profil en termes de types de réponse). Nous réalisons ainsi un certain nombre d'inférences relatives à la fois à l'identité professionnelle des répondants et à leur volonté d'action et/ou de changement.

## 4. Analyse des questionnaires et résultats

Les réponses des enseignants au questionnaire sont mises en perspective et illustrées par des *verbatim* d'entretiens. Nous conservons pour cette présentation l'ordre des rubriques structurant le questionnaire.

### 4.1. Caractéristiques individuelles de la population interrogée par questionnaire

Les 81 répondants (60 hommes, 21 femmes) sont des enseignants-chercheurs de l'UFR de physique de l'université Paris-Diderot ; ils se répartissent de façon à peu près homogène si l'on considère leur nombre d'années d'expérience en tant qu'enseignant. Rappelons-le, les professeurs des universités répondent davantage que les maîtres de conférences et les moniteurs (73 % des professeurs et 29 % des maîtres de conférences de l'UFR répondent au questionnaire).

98 % des EC interrogés déclarent avoir appris à enseigner au fil de leur expérience professionnelle. Certains d'entre eux (près d'un quart) mentionnent la préparation à l'agrégation, élément que nous avions retenu suite aux remarques de plusieurs EC lors des entretiens. Toujours à propos de la formation à l'enseignement, la moitié des EC ayant répondu au questionnaire évoque, en plus de son expérience, les échanges avec les collègues. Précisons également que les EC interrogés valorisent autant la recherche que l'enseignement lorsqu'on leur demande « ce qui leur plaît le plus » dans leur métier d'enseignant-chercheur. Parmi les EC interrogés, personne (à une exception près) n'affirme ne pas savoir enseigner.

Les deux questions portant sur différents aspects de l'activité pédagogique d'un EC susceptibles d'être (ou non) délégués (c'est-à-dire : faits par d'autres qu'eux-mêmes) nous ont permis d'approcher d'autres éléments spécifiques de l'identité professionnelle des enseignants-chercheurs interrogés (en termes d'appétence/de rejet déclarés). Les résultats présentés ci-après (figure 1) montrent que les enseignants-chercheurs valorisent en priorité l'enseignement au niveau master et qu'ils sont attachés à l'enseignement en présentiel (seuls cinq EC sont prêts à le déléguer). *A contrario,* les tâches de contrôle (colles, corrections) sont les tâches les moins valorisées. Une grande majorité est prête à les déléguer. Concernant l'accueil et les contacts individualisés avec les étudiants, les résultats sont partagés. Nous verrons pourtant par la suite l'importance que tous accordent à cette fonction d'accompagnement personnalisé. Cette légère contradiction est sans doute liée à la tension qui existe entre ce que les enseignants-chercheurs apprécient en termes de pratique du métier, et ce qu'ils considèrent comme efficace ou important pour les étudiants (objectifs du métier) ; deux considérations qui ne concordent pas forcément. Cette non-concordance explique sans doute d'autres tensions que nous signalerons plus avant.

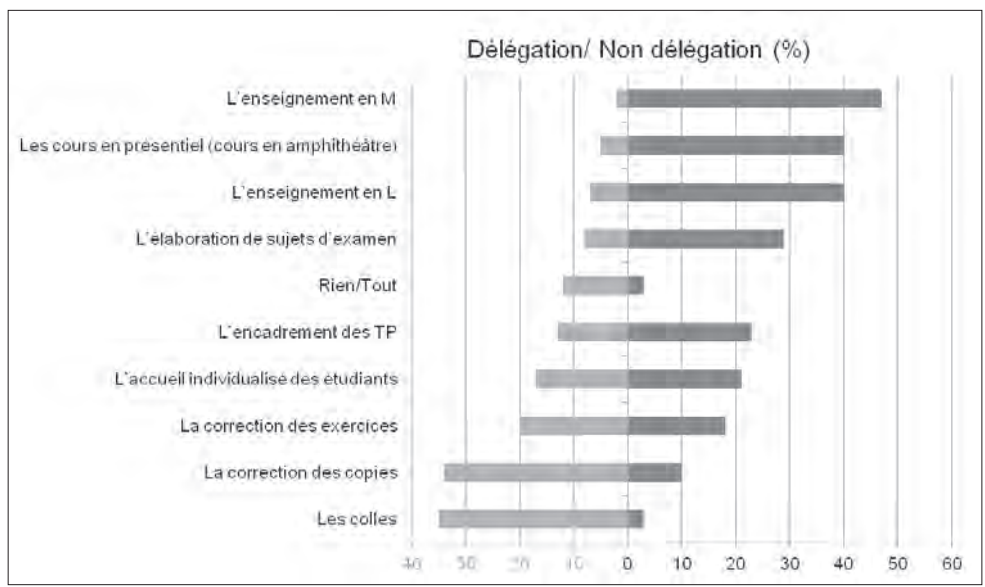

Fig. 1 : graphique présentant les pourcentages d'EC attachés à la délégation ou à la non-délégation des tâches requises par leur métier d'enseignant.

### 4.2. L'organisation pédagogique des enseignements

Conformément aux résultats des enquêtes précédentes, la passivité des étudiants demeure l'objet de préoccupation principal des enseignants-chercheurs (notamment lorsqu'on les interroge en entretien) et semble former le fil conducteur de leur propos (la passivité est souvent mise en relation avec le manque d'autonomie, de travail personnel, de participation). Pour la plupart d'entre eux, les pratiques effectives d'enseignement et les modalités au sein desquelles elles se vivent ne permettent pas de surmonter cette difficulté.

### *4.2.1. L'amphi, une modalité irremplaçable qui doit favoriser l'activité des étudiants*

Si l'on entre dans le détail des modalités d'organisation pédagogique, les enseignants-chercheurs accordent une importance majeure au cours en amphithéâtre ; 61 % répondent que cette modalité n'est remplaçable par rien d'autre. Rappelons également que c'est une modalité qu'ils ne sont pas prêts à déléguer. L'amphi est également perçu par certains EC comme un espace au sein duquel le savoir académique est présenté de façon à ce que l'étudiant puisse « prendre de la hauteur », un espace qui permet « d'aller un peu plus loin » :

> *« Disons que j'ai rien contre un cours magistral qui prenne un peu de hauteur. Au contraire, je pense qu'il faut tirer vers le haut. On entend parler des cours-TD. Un cours-TD qui colle presque au TD avec une application directe à des fondamentaux de cours, comme on peut voir dans un certain nombre de bouquins, Hachette, etc., c'est utile, mais il faut aller un peu plus loin je pense et ça peut se faire en cours. »* [enseignant-chercheur E12]

Pourtant, plus de la moitié des enseignants (58 %) ayant répondu que la modalité « amphi » n'est remplaçable par rien d'autre approuvent un passage en « cours-TD » en tant que modalité exclusive d'enseignement. Pour ces enseignants, c'est certainement le contenu de l'espace « amphi » qui est à modifier ; en d'autres termes, il faut maintenir la modalité « amphi » (en tant qu'espace d'enseignement devant un grand groupe) et changer ce que l'on y fait :

> *« L'amphi ça fonctionne, on arrive à former certains étudiants comme ça. En tout cas, on essaie […]. Il faut continuer à faire un cours qui reste un cours d'amphi. C'est-à-dire/on peut être nombreux mais on peut changer, faire peut-être plus d'exemples, plus d'exercices, des manips de cours pour que ce soit plus interactif. »* [enseignant-chercheur E20]

D'ailleurs, « l'amphi » n'est pas perçu comme un espace d'enseignement dans lequel l'étudiant est censé être passif. En effet, la participation (67 %) et la présence (50 %) sont les éléments qui permettent aux EC interrogés de savoir si l'objectif du cours en amphi a été atteint tout en regrettant que cela ne soit pas davantage le cas dans les faits :

> *« Il faudrait peut-être avoir une approche un peu plus interactive au niveau des cours plutôt que d'avoir quelque chose du type : on se pose, on fait l'amphi, ça se termine et on s'en va. »* [enseignant-chercheur E11]

### *4.2.2. Le TD, une modalité à modifier en profondeur pour favoriser l'autonomie des étudiants*

Les EC interrogés affichent un rejet plus franc en direction de la modalité « TD », en tout cas, dans sa pratique effective, rejet que nous avions largement constaté au cours des entretiens exploratoires. Moins d'un tiers des enseignants ayant répondu au questionnaire affirment que cette modalité n'est remplaçable par rien d'autre et près de 60 % semblent favorables au fait que les TD soient remplacés par du tutorat.

L'analyse des réponses des EC vis-à-vis des TD laisse à nouveau penser qu'il existe un *hiatus* entre ce qu'ils y font (pratique du métier) et ce qu'ils y projettent (objectifs du métier), projection à laquelle les entretiens exploratoires nous permettent d'associer les attributs suivants :

- Un TD « efficace » est un TD qui a été préparé par l'étudiant ;
- Le rôle de l'enseignant ne doit pas se limiter à de la correction d'exercices au tableau, il doit être « médiateur » (ceci est à mettre en lien avec le fait que le tutorat est plébiscité) ;
- Un TD doit être un « espace d'interactions ».

On retrouve ici l'idée selon laquelle le TD tel qu'il devrait être, tel qu'il a été pensé, n'est pas inapproprié, mais tel qu'il est pratiqué actuellement dans son espace de contrainte, sa forme est insatisfaisante.

### 4.2.3. Le cours-TD, une modalité alternative

La modalité « cours-TD » est perçue comme une voie permettant de pallier les carences de la modalité « amphi » en termes d'interactions enseignant-étudiants. Finalement, 65 % des EC interrogés approuvent une organisation qui repose sur la mise de « cours-TD » en tant que modalité exclusive :

> *« Je suis un peu mal à l'aise avec les cours d'amphi parce que j'ai vraiment l'impression qu'il y a peu d'étudiants qui en tirent profit et puis les enseignants ils refont des cours qui sont déjà dans les bouquins. En plus il y a peu d'interactions, les étudiants ne posent pas de question. Je pense qu'il faudrait plutôt des séances qui ressemblent plutôt à des cours-TD, on leur donne un chapitre à lire, ils posent toutes les questions qu'ils veulent sur un chapitre qu'ils ont dû lire et on fait une espèce de cours-TD où on revoit les points des chapitres et on fait des exercices d'application. »* [enseignant-chercheur E5]

Les réserves (35 % des EC interrogés sont défavorables à cette modalité, en tant que modalité exclusive d'enseignement) sont d'ordre technique (ex : les moyens – en heures – nécessaires, le nombre de salles, etc.) et pédagogique (un fonctionnement en cours-TD rapprocherait l'enseignement universitaire de l'enseignement secondaire).

Enfin, il apparaît fortement consensuel de maintenir la possibilité d'un contact (privilégié) avec les étudiants : 90 % des EC sont en effet favorables à la tenue d'une « permanence d'accueil », même si, rappelons-le, beaucoup se disent prêts à la déléguer (voir figure 1).

### 4.3. Les difficultés associées à l'enseignement de la physique

69 % des EC interrogés trouvent que le trop grand nombre d'heures d'enseignement constitue la principale difficulté du métier d'enseignant. Cette dimension demeure prégnante dans les entretiens ; elle est souvent mise en relation avec le temps requis pour l'activité de recherche.

Seuls 18 % des EC interrogés choisissent le manque de formation pédagogique en tant que difficulté pour la pratique du métier d'enseignant de physique. Nous reviendrons sur ce dernier point ultérieurement.

Concernant les difficultés des étudiants (telles que perçues par les enseignants-chercheurs interrogés, voir figure 2), le manque de travail est la raison principalement plébiscitée. Dans les entretiens, cette dimension est souvent mise en lien avec le manque de temps disponible :

*« Il faut être clair, les étudiants aujourd'hui on ne les lâche jamais. Ils sont ici de 8 h du matin à 18 h le soir avec souvent 1 h à 1 h 30 de trajet par jour. Il ne faut pas rêver, dans ces conditions, ils ne peuvent pas travailler chez eux. »* [enseignant-chercheur E1].

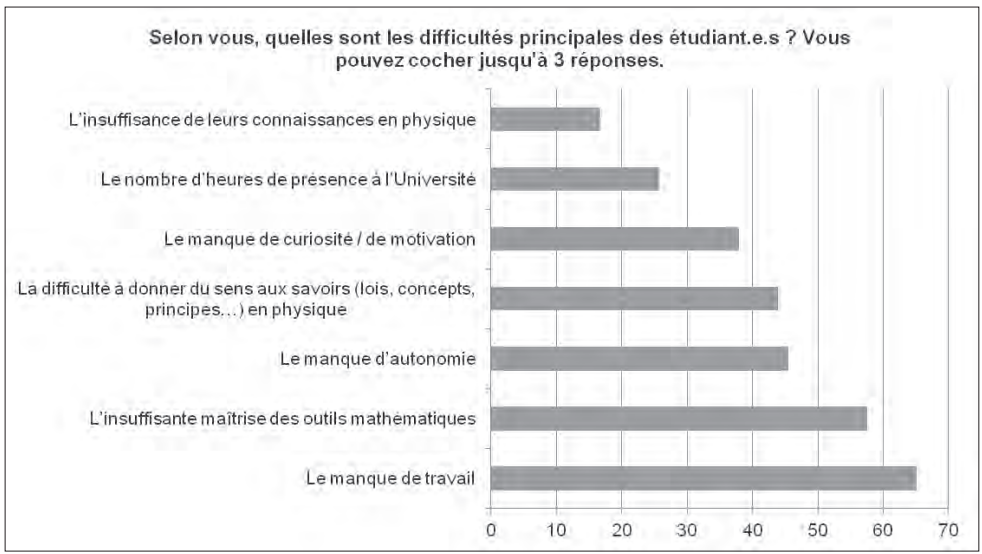

**Fig. 2 : répartition du nombre de réponses des EC auxquels on demande de choisir 3 difficultés d'étudiant au sein d'une liste imposée.**

Plus de la moitié des EC ayant répondu au questionnaire (57 %) déplorent l'insuffisante maîtrise des outils mathématiques des étudiants. Cet aspect est souvent considéré (dans les entretiens) comme un obstacle à donner à l'outil mathématique son sens « physique » :

*« Une des grosses grosses difficultés c'est le calcul avec des petites variations. Pour eux c'est très abstrait et cette abstraction n'est pas seulement mathématique, c'est un peu tout : c'est avoir compris ce que font les physiciens quand ils font ces choses-là. »* [enseignant-chercheur E14]

Pour ces enseignants-chercheurs, ces écueils sont légitimes et étroitement liés à la difficulté intrinsèque du savoir en jeu (notamment mathématique) dont les étudiants peinent à prendre la mesure (notamment en terme de quantité de travail, d'effort à fournir) :

*« Moi je trouve que les élèves qui arrivent ne sont pas habitués à transpirer pour apprendre. Il faut développer le goût de l'effort. Les étudiants ne savent pas que c'est formel, qu'on doit transpirer, que c'est pas simple. La physique avec les mains c'est bien pour les enfants de 10 ans, et encore. Ça donne*

*de mauvaises habitudes. Ils ont du mal avec le formalisme, les mathématiques et avec la conceptualisation.* » [enseignant-chercheur E2]

Au-delà de l'outil mathématique lui-même, c'est la capacité à donner du sens au savoir (lois, concepts, principes, etc.) en physique qui est mentionnée à plusieurs reprises au cours des entretiens et considérée comme une difficulté importante pour 43 % des EC interrogés :

*« Souvent c'est l'acuité du raisonnement qui ne va pas Il y a un flou dans le raisonnement des étudiants […]. Ce serait un maniement approximatif des concepts, ils confondent force et énergie cinétique par exemple. C'est un problème d'expression, écrite ou orale, qui traduit largement la rigueur et l'organisation de ta propre pensée. »* [enseignant-chercheur E1]

Le manque de curiosité, choisi par 38 % des EC interrogés, conduit parfois les enseignants-chercheurs que nous avons interviewés à proposer des formes alternatives de pratiques pédagogiques de façon à susciter la curiosité des étudiants :

*« Si tu présentes une expérience dont le résultat est un peu inattendu tu peux créer une rencontre avec les étudiants et c'est ça au fond qui est important, c'est la rencontre. Parce que quand tu fais une expérience tu t'attends à un résultat et t'en obtiens un autre, alors soit tu t'es planté dans la théorie, soit autre chose et là avec les étudiants tu dis « ben tiens, qu'est-ce qui se passe ? » ben là ça devient marrant. »* [enseignant-chercheur E6]

Ce faisant, c'est finalement leur propre pratique qu'ils engagent en mettant en perspective les difficultés des étudiants et leurs propres modalités d'intervention, ce qu'illustrent les extraits suivants :

*« Les difficultés des étudiants, pour moi…. Ben c'est les mauvais enseignants, non ? »* [enseignant-chercheur E10]

*« Pour moi, il y a deux sortes de difficultés : les difficultés techniques […] et les difficultés conceptuelles. C'est compliqué la physique et je suis effrayée par la vitesse à laquelle ça va […]. T'as pas le temps de comprendre […]. À la fois, c'est peut-être moi qui me trompe d'objectif. Peut-être qu'on leur demande pas de comprendre la physique, on leur demande d'assimiler des connaissances, de faire des calculs mais on leur demande pas d'avoir compris la physique, on teste autre chose. »* [enseignant-chercheur E3]

*« Dire "ils savent pas faire de maths", ça c'est vrai, mais nous on connaissait pas l'algèbre symplectique et pourtant on a fait de la physique. Alors dire "les élèves connaissent pas le produit scalaire", bon, c'est à nous de nous adapter. On a les gens qu'on a et il faut qu'on travaille à partir de ce public là. »* [enseignant-chercheur E5]

Soulignons enfin que l'insuffisance des connaissances en physique des étudiants ne semble pas constituer une difficulté particulière pour les EC interrogés.

### 4.4. L'évaluation (des étudiants, des enseignements)

89 % des EC interrogés sont favorables au maintien d'un examen final sans pour autant plébisciter cette modalité d'évaluation en tant que modalité exclusive (pas plus qu'une modalité exclusive de contrôle continu). La modalité qui remporte le maximum de suffrages est une modalité mixte de contrôle terminal (examen final) et de contrôle continu (qui, dans la plupart des réponses, devrait prendre la forme de devoirs à la maison). Certains enseignants-chercheurs mentionnent le fait que la mise en place d'un contrôle continu systématique induit une charge supplémentaire de travail (préparation, corrections de copies), charge qui, selon eux, devrait faire partie intégrante du service d'enseignement.

Toujours à propos de l'évaluation des étudiants, plusieurs enseignants-chercheurs regrettent (propos recueillis lors des entretiens) que l'évaluation des étudiants soit considérée à la fois par les étudiants et par les enseignants comme une sanction :

*« L'évaluation aujourd'hui c'est reproduire en un temps limité l'exercice du TD. Et c'est une sanction et ça n'aide pas. Déjà si on s'intéresse à la formation des étudiants, l'évaluation ne doit pas être une sanction. »* [enseignant-chercheur E9]

Certains remarquent également que le système universitaire ne favorise pas forcément l'adéquation entre « apprentissage » et « réussite » :

*« Apprendre des choses et maximiser ses chances de réussite sont deux choses différentes et il faut essayer de faire des règles d'évaluation qui font que ces deux choses-là se recoupent le plus précisément possible. »* [enseignant-chercheur E11]

Il est d'ailleurs intéressant de remarquer que la plupart des EC interrogés ne choisissent pas « la réussite au partiel ou à l'examen » pour apprécier si l'objectif de leur enseignement en amphi a été atteint mais bien la participation des étudiants. Or, cette participation n'est pas un objet d'évaluation (cela pourrait être cependant le cas si la culture universitaire reconnaissait la valeur de l'évaluation formative). Seule est évaluée la capacité à résoudre des problèmes, capacité à laquelle forment les TD.

Parce qu'ils participent de l'évaluation (davantage formative) des étudiants, nous avons souhaité interroger les enseignants-chercheurs sur la pertinence des systèmes interactifs de vote électronique (*clickers*) dont l'usage, appuyé par la recherche en didactique de la physique (Smith *et al.*, 2009 ; Deslauriers, Schelew & Wieman, 2011)[10] commence à se répandre, notamment dans les universités nord-américaines. Notons également que les *clickers* ont été mentionnés spontanément dans plusieurs entretiens en tant que pratique innovante[11].

---

10　Ces recherches évaluent l'impact de *clickers* en grands groupes d'enseignement associés à des questions à choix multiples construites en référence aux conceptions des étudiants. Les réponses aux questions (qui touchent à différents domaines de la physique) sont proposées par les étudiants après discussion collective en petits groupes. La confrontation à la réponse correcte (souvent non conforme à la réponse des étudiants) suscite la curiosité et développe une forme d'appétence pour l'apprentissage.

11　Précisons sur ce point que la plupart des enseignants-chercheurs que nous avons interrogés disent ne pas avoir mis en place de pratiques qui pourraient être qualifiées d'innovantes (l'attribut est d'ailleurs moqué à plusieurs reprise). Certains évoquent toutefois l'usage du vidéoprojecteur ou la mise en place d'interrogations régulières en début de TD. Les *clickers* sont évoqués à plusieurs reprises en tant qu'innovation possible par des enseignants plutôt enclins à une généralisation de la pédagogie dite « inversée » *via*, par exemple, les MOOC (Minichiellio, 2013).

Moins de la moitié des EC interrogés (44 %) se déclarent favorables à leur usage. Si certains d'entre eux y voient une façon d'accéder aux difficultés de compréhension en temps réel et un moteur pour la mise en activité collective des étudiants dans le cadre d'un enseignement en amphi, d'autres, considèrent qu'il s'agit d'un « gadget ridicule », à l'image du commentaire suivant (accompagnant une réponse « non » à la question « seriez-vous favorable à l'usage des *clickers* ») :

> « Il n'y a aucun intérêt à connaître "en temps réel" si un étudiant a compris un enseignement : comprendre, c'est s'approprier une connaissance et cela ne se fait pas un instant. » [commentaire au questionnaire]

Concernant les enseignements, 95 % des EC interrogés sont favorables à leur évaluation[12]. La majorité d'entre eux plébiscitent une modalité « externe » d'évaluation (questionnaires distribués aux étudiants par le responsable de la spécialité ou par le responsable de l'UE) souvent associée à une « discussion collective » avec les étudiants (42 %). 75 % souhaitent cette évaluation à l'issue de l'enseignement ce qui laisse penser que l'évaluation là encore n'est pas perçue dans sa fonction formative (une évaluation des enseignements en cours de semestre permettrait un réajustement des orientations didactiques et pédagogiques choisies par l'enseignant). Certains enseignants-chercheurs (17 %) déplorent enfin le fait que l'activité d'enseignement ne soit pas évaluée au niveau institutionnel au même titre que l'activité de recherche.

### 4.5. La formation des enseignants-chercheurs

La question de la formation des enseignants-chercheurs est certainement celle qui est apparue la moins consensuelle. La moitié des EC interrogés y est favorable, l'autre non ; et les raisons du rejet de l'idée de formation peuvent être résumées par les trois phrases suivantes : « je n'ai pas le temps pour cela », », « je connais le système éducatif de l'intérieur donc je sais ce qu'est un bon enseignant », « je suis un bon chercheur donc je suis un bon enseignant » :

> « Bon, déjà pour être enseignant-chercheur, il faut être capable de se présenter, de présenter des idées puisque c'est le passage obligé du recrutement. Donc en général, ceux qu'on recrute, ils savent enseigner, en tout cas, ils savent communiquer et faire passer des messages. » [enseignant-chercheur E6]

Au cours des entretiens, la question de la formation continue des EC avait paru recueillir davantage les faveurs des enseignants que nous avions interrogés. Sur ce point (et comme nous le signalions au début de cet article), certains avaient plébiscité la préparation à l'agrégation. Pour ces enseignants, la nécessité de la formation était perçue comme un besoin de remise à niveau académique. D'autres avaient évoqué des besoins d'ordre pédagogique, voire didactique auxquels pourraient répondre des séminaires d'échanges de pratiques :

> « J'aimerais pouvoir discuter de pratique pédagogique avec des collègues. On fait bien des séminaires de recherche, pourquoi on ferait pas des séminaires sur l'enseignement, des tables rondes ? […] Moi ça me terrorise par exemple de faire un amphi alors je suis même preneur d'une formation "comment faire un amphi en histoire de l'art" parce que je suis sûr qu'il y a des choses qui se transposent. » [enseignant-chercheur E7]

> « La formation continue d'un enseignant-chercheur ça pourrait être un truc comme ça : soit sur un point très précis de physique, les difficultés que risquent d'avoir les étudiants, soit la gestion d'un groupe, par exemple pour gérer un groupe de L1 qui veut pas bosser. » [enseignant-chercheur E3]

---

12 Il est important de préciser que la question ne laissait aucune autre possibilité d'évaluation des enseignements que le recours à l'opinion des étudiants, ce qui peut paraître tout à fait restrictif. À ce propos, un enseignants interviewé signale : « il n'y a pas plus cruel que la correction des copies ; une copie c'est bien plus cruel que l'avis d'un étudiant » [enseignant-chercheur E13].

55 % des EC interrogés accepteraient d'être filmés pendant leur cours ou leur TD et de participer à une (ou plusieurs) séance d'autoconfrontation (Yvon & Garon, 2006)[13]. Lors des entretiens, certains enseignants avaient spontanément évoqué l'intérêt qu'ils verraient, en termes de formation, à se voir en train de faire cours :

> « Ça serait vachement bien un truc genre "autoscopie" en plus c'est simple à mettre en place. La question c'est "juger pour s'améliorer" ou "juger pour être puni". Dans le système scolaire, les notes sont perçues comme un moyen de "punir" alors que ça devrait être un moyen de s'améliorer. La notation pour la sélection est mal vécue par les EC qui la refusent. Donc l'idée d'une autoscopie ce serait vraiment intéressant. Et faire un retour d'expérience "moi je l'ai fait, ça a donné ça… " » [enseignant-chercheur E5]

Ainsi, lorsque l'enseignant-chercheur accepte d'être formé, c'est plutôt de manière réflexive ce que soulignaient déjà Donnay et Romainville en 1996.

## 5. Synthèse des résultats et discussion

Ce travail d'enquête est né d'une commande d'une UFR de physique. Il s'agissait de faire un état des lieux des différents positionnements des enseignants-chercheurs de cette UFR face aux dispositifs organisationnels (CM, TD, etc.), aux modalités d'évaluation, de formation, aux facteurs favorisant/bloquant la réussite des étudiants, etc., afin de soutenir un projet de rénovation de l'organisation et des méthodes pédagogiques en place. Au-delà des propositions concrètes qui ont pu émerger, cette enquête a permis, dans sa dimension heuristique, d'apporter plusieurs éléments de connaissance sur les conceptions que les enseignants-chercheurs interrogés ont de l'enseignement, et de dégager de ce fait quelques marqueurs de leur identité professionnelle (considérée du point de vue de leur activité d'enseignant).

---

13   98 % des EC ressentant le besoin d'être formés se déclarent intéressés par une ou plusieurs séances d'auto-confrontation.

*5.1. Un bon enseignant est avant tout un bon physicien*

L'identité professionnelle de l'enseignant-chercheur, dans son métier d'enseignant, est marquée par son attachement au savoir de sa discipline : la physique. Cette affirmation est soutenue par les indices suivants :

– Un attachement à l'espace magistral d'exposition du savoir. Ce point avait déjà été souligné par Altet il y a dix ans : « certains universitaires affirment aimer les grands amphis, ils ont du plaisir à y enseigner et à faire du cours magistral » (Altet, 2004, p. 42) ;

– Une valorisation de l'enseignement au niveau master (espace au sein duquel le savoir enseigné est souvent la cible scientifique des recherches des enseignants-chercheurs qui y enseignent) ;

– Une valorisation de l'activité de communication de la recherche (sous-tendue par l'idée que cette activité est directement transposable à une situation d'enseignement, ce qui n'a pourtant rien d'évident) ;

– Une valorisation de l'agrégation en tant que modalité de formation ;

– Une minoration de l'insuffisance des connaissances en physique des étudiants ; la transmission du savoir en physique semblant ainsi relever de la pleine responsabilité de l'enseignant-chercheur.

Il nous semble important de préciser qu'affirmer qu'« un bon enseignant est avant tout un bon physicien » positionne l'enseignant-chercheur d'une manière qui diffère quelque peu de l'affirmation selon laquelle « les enseignants chercheurs sont physiciens avant d'être enseignants de physique » (Endrizzi, 2011). Ici, les qualités pour être enseignant semblent puisées dans le rapport au savoir disciplinaire sans qu'il n'apparaisse de phénomène de prédominance d'un métier (la recherche) sur l'autre (l'enseignement). Dit autrement, tout se passe comme si l'enseignant-chercheur puisait dans son identité de chercheur les ressources pour bien enseigner. D'ailleurs, l'activité d'enseignement apparaît autant valorisée que l'activité de recherche et l'attachement au savoir disciplinaire n'est pas exclusif d'un intérêt marqué pour la recherche d'une plus grande interactivité avec les étudiants.

*5.2. L'enseignement doit favoriser l'interaction avec les étudiants*

Ce qui guide souvent les propositions de changement énoncées par les enseignants-chercheurs que nous avons interrogés semble plutôt associé au combat à mener contre la passivité des étudiants[14] (maintes fois soulignée) et à la difficulté constatée de favoriser leur mise au travail (c'est d'ailleurs le « manque de travail » qui constitue la plus grande source de difficulté des étudiants du point de vue des EC interrogés). Il se dégage donc, à l'issue de cette enquête, un consensus sur le fait d'inciter les étudiants à travailler par eux-mêmes (quitte à alléger les horaires de présence à l'université, le contenu des feuilles d'exercices, ou à utiliser un ouvrage unique), et sur l'idée de favoriser les interactions étudiants-étudiants et étudiants-enseignant (via un système de tutorat, ou au sein même de l'espace « amphithéâtre »). On notera à ce propos que les modalités d'intervention « à distance » (cours, exercices en lignes) sont très peu valorisées en tant que modalités alternatives d'enseignement, la modalité présentielle restant pleinement plébiscitée. La recherche d'interactivité, notamment en amphithéâtre, avait été relevée par Clanet (2004). Malheureusement, les stratégies mises en place ou citées comme fécondes par les enseignants-chercheurs (jeu de questions-réponses, appel à des situations de la vie courante, etc.) ne semblaient pas repérées par les étudiants.

---

14. Une enquête plus approfondie permettrait d'ailleurs de savoir ce que les enseignants-chercheurs entendent par « passivité », le terme n'ayant jamais été défini lors des entretiens, un peu comme si cela allait de soi qu'un étudiant « qui ne fait rien », « qui ne bouge pas » est un étudiant qui n'apprend pas ; comme si, au contraire, apprendre nécessitait d'être visiblement « actif ».

Finalement, il semble que les EC que nous avons interrogés présentent des traits identitaires communs avec deux types d'enseignants que Kember (1997) semblait distinguer : les enseignants dits « magio-centrés » (c'est-à-dire centrés sur eux-mêmes et sur le contenu disciplinaire), les enseignants centrés sur les « interactions étudiants-enseignant ». Cette séparation semble peu appropriée ici. Les EC interrogés inscrivent leur métier d'enseignant à la fois dans le rapport enseignant-savoir et dans le rapport enseignant-étudiants.

### 5.3 Enseigner la physique à l'université : un métier en tension

L'identité professionnelle des EC de physique que nous avons interrogés apparaît fortement marquée par des tensions (Altet, 2004), tensions qui se révèlent parfois sous la forme « je sais qu'il faudrait faire ceci et pourtant je fais le contraire ». Ainsi, s'ils affichent une réelle volonté de changement accompagnée le plus souvent de propositions concrètes (suppression des TD, passage au cours-TD, généralisation du contrôle continu, utilisation des outils de sondages électroniques, autoconfrontation, etc.), les EC semblent contraints par des éléments de difficulté qu'ils expriment assez clairement et qui relèvent de la perception qu'ils ont d'eux-mêmes (chercheur qui reste attaché aux lieux d'expression magistrale du savoir académique, démuni face à la passivité des étudiants), de l'institution (trop d'heures d'enseignement, activité professionnelle non évaluée, non valorisée, à laquelle on se forme « sur le tas ») et des étudiants (passifs, qui ne travaillent pas suffisamment et qui peinent à conceptualiser, à donner du sens aux concepts, lois, principes qui fondent la physique).

Leurs propos sur les règles liées aux dispositifs organisationnels sont, à ce titre, tout à fait éclairants. Les TD font l'objet d'un rejet massif et pourtant, la façon dont la plupart des enseignants-chercheurs que nous avons interrogés s'assure que l'objectif d'un TD (c'est-à-dire faire des exercices) a été atteint reste la réussite à l'examen. Au contraire, l'amphithéâtre demeure plébiscité mais la façon dont on s'assure que son objectif a été atteint est directement liée à la participation des étudiants et à leur présence. Autrement dit, ce que l'on cherche à évaluer (ce qui est étudié en TD) n'est pas ce qui est valorisé ou considéré comme irremplaçable (le cours en amphi).

## Conclusion

Les changements dans les pratiques pédagogiques à l'université sont souvent le fait d'un petit nombre d'enseignants et les innovations peinent à investir d'autres espaces que ceux de leurs promoteurs. L'usage des technologies numériques est, à ce titre, exemplaire. Valluy (2013) signale à cet égard que les processus de construction des argumentaires en faveur des TICE à l'université excluent la plupart du temps les points de vue critiques. Pour Valluy, les acteurs de ces processus « ont l'impression de débats ouverts à une diversité de points de vue alors qu'ils se confortent dans leurs convictions par des consensus endogènes excluant d'autres regards, plus critiques ». Il poursuit en précisant que l'éviction de ces points de vue condamne à se priver de leurs apports. Ainsi, faute d'en tenir compte, les promoteurs des TICE prennent le risque « de propulser des décisions technologiques et politiques inadaptées, rencontrant des échecs de mise en œuvre en raison même des évictions initiales » (Valluy, 2013).

Les propositions qui ont finalement émergé de notre enquête (mise en place de « manips » de cours en amphithéâtre, installation de groupes de travail autonomes en TD, mise en place d'interrogations écrites en début de TD, usage d'un ouvrage unique), ont été testées dès la rentrée 2013 au sein de l'ensemble des enseignements de physique du L1 de sciences exactes dans un relatif consensus. Dans notre démarche, la prise en compte d'éléments d'identité professionnelle des enseignants-chercheurs interrogés a constitué une condition pour que l'innovation puisse prendre corps de manière consensuelle, au-delà des désirs et des intimes convictions de quelques-uns. Cette recherche de consensus a d'ailleurs orienté notre travail puisqu'en recherchant les traits communs de l'identité professionnelle des EC que nous avons interrogés nous avons mis de côté l'analyse des variations d'un enseignant à l'autre.

À ce propos, il est intéressant de remarquer que, même s'ils sont rares, certains EC évoquent de manière spontanée des aspects directement liés à la nature de la physique et/ou aux difficultés conceptuelles des étudiants, en physique. Interrogés sur ce qui, de leur point de vue, constitue des difficultés pour les étudiants, certains évoquent (lors des entretiens, notamment) le caractère intrinsèquement « difficile » de la physique. Difficile parce que mobilisant un formalisme de nature mathématique ; difficile aussi parce que nécessitant une grande capacité d'abstraction, les mots, la langue devant prendre sens dans un système conceptuel distinct de celui du « sens commun ». D'autres voient dans la physique une occasion de balayer les idées reçues en mettant à profit son potentiel à générer des situations expérimentales « inattendues ». Au cours des entretiens, deux EC nous ont également demandé s'il existait des « banques de nœuds conceptuels », des « listes de difficultés récurrentes en mécanique, en optique, etc. » (aspects qui n'ont pas été très valorisés dans les réponses au questionnaire).

Dans une perspective d'exploration plus spécifiquement « didactique », l'une des suites de ce travail vise à mieux comprendre l'identité professionnelle des enseignants-chercheurs dans leur rapport spécifique à la physique, et à identifier les conséquences de ce rapport sur leurs choix d'enseignement, en termes de contenus et de déroulements (Robert, 2003). Les premières observations de pratiques d'enseignement conduites *in situ* pourraient permettre d'éclairer les questions suivantes : quels sont les marqueurs de la culture épistémologique et didactique de l'enseignant-chercheur lorsqu'il enseigne ? Comment un enseignant-chercheur averti des difficultés récurrentes des étudiants (dans un domaine spécifique de la physique) adapte-t-il son enseignement ? Quelles sont les conséquences de ses choix sur l'apprentissage des étudiants en physique ? Les réponses à ces questions pourront, à terme, alimenter des formations pour des enseignants-chercheurs de physique, et être orientées vers la mise en place effective d'un système au sein duquel enseignement et apprentissage sont étroitement imbriqués.

**Cécile de Hosson**
cecile.dehosson@univ-paris-diderot.fr

**Nicolas Décamp**
nicolas.decamp@univ-paris-diderot.fr

**Émilie Morand**
emorand@stravia.net

**Aline Robert**
aline.robert@u-cergy.fr

## Bibliographie

# Annexe

# Questionnaire

*Informations générales*

Je suis
- ☐ Un homme
- ☐ Une femme

Nombre d'années d'expériences d'enseignement
- ☐ Moins de 5 ans
- ☐ Entre 5 et 15 ans
- ☐ Plus de 15 ans

Corps
- ☐ Doctorant.e
- ☐ MCF
- ☐ PR

Indiquez votre parcours de formation. Vous pouvez cocher plusieurs cases.
- ☐ Classes Préparatoires aux Grandes Écoles
- ☐ École d'ingénieurs
- ☐ École Normale Supérieure
- ☐ Université à partir du premier cycle
- ☐ Université à partir du second cycle
- ☐ Université à partir du troisième cycle
- ☐ Agrégation
- ☐ Formation supérieure à l'étranger, précisez
- ☐ Autre, précisez

*Identité professionnelle*

Classez par ordre de préférence (de 1 à 3) ce qui vous plait le plus dans votre métier d'enseignant.e-chercheur.e ?
- ☐ L'équilibre entre enseignement et recherche
- ☐ L'enseignement
- ☐ Le travail en équipe
- ☐ La recherche
- ☐ L'écriture d'article
- ☐ Les colloques

☐ Les tâches administratives
☐ La diversité des activités professionnelles
☐ Autre, précisez

Comment avez-vous appris à enseigner ?
☐ Au fil de mon expérience
☐ En passant l'agrégation
☐ Je ne sais pas enseigner
☐ En lisant des ouvrages de pédagogie universitaire
☐ En suivant des formations
☐ En échangeant avec les collègues
☐ Autre, précisez

Que seriez-vous prêt à déléguer en tant qu'enseignant.e ?
☐ La correction des copies
☐ L'élaboration de sujets d'examen
☐ L'enseignement en L
☐ L'enseignement en M
☐ L'accueil individualisé des étudiants
☐ Les cours en présentiel (cours en amphithéâtre)
☐ La correction des exercices
☐ L'encadrement des TP
☐ Les colles
☐ Aucune, je ne souhaite rien déléguer
☐ Autres, précisez

*A contrario*, quelles sont les pratiques parmi cette liste que vous ne délégueriez pas ou que vous n'aimeriez pas devoir déléguer ?
☐ La correction des copies
☐ L'élaboration de sujets d'examen
☐ L'enseignement en L
☐ L'enseignement en M
☐ L'accueil individualisé des étudiants
☐ Les cours en présentiel (cours en amphithéâtre)
☐ La correction des exercices
☐ L'encadrement des TP
☐ Les colles
☐ Aucune, je suis prêt.e à tout déléguer
☐ Autres, précisez

**L'organisation pédagogique des enseignements**

Quels sont, selon vous, les objectifs d'un cours en amphithéâtre ? En cas de réponses multiples, numérotez-les trois objectifs principaux selon leur ordre d'importance.

- ☐ Apport de connaissances en physique (lois, concepts, principes)
- ☐ Apport de techniques de résolution de problème
- ☐ Explicitation de démonstrations
- ☐ Mise en contexte culturel (historique, par exemple) des connaissances en physique
- ☐ Autre, précisez

Comment vous assurez-vous que cet objectif a été atteint ?
- ☐ Participation des étudiants
- ☐ Présence au cours
- ☐ Poursuite des études en physique
- ☐ Réussite à l'examen / au partiel
- ☐ Implication des étudiant.e.s dans la résolution des exercices en TD
- ☐ Sollicitation des enseignant.e.s par les étudiant.e.s (par mail, en TD, dans son bureau, etc.)
- ☐ Autre, précisez

La modalité « cours en amphithéâtre » vous semble-t-elle remplaçable par…
- ☐ Une vidéo accessible *via* l'intranet
- ☐ Un ouvrage
- ☐ Un poly
- ☐ Les cours en amphi ne sont remplaçables par rien d'autre
- ☐ Autre, précisez

Quels sont, selon vous, les objectifs d'un TD ? En cas de réponses multiples, numérotez-les trois objectifs principaux selon leur ordre d'importance.
- ☐ Apport de connaissances en physique (lois, concepts, principes)
- ☐ Apport de techniques de résolution de problème
- ☐ Explicitation de démonstrations
- ☐ Correction d'exercices
- ☐ Application des lois théoriques vues en amphi
- ☐ Préparation des étudiant.e.s à la réussite de l'examen
- ☐ Apport de réponses à des questions d'étudiant.e.s
- ☐ Autre, précisez

Comment vous assurez-vous que cet objectif a été atteint ?
- ☐ Participation des étudiants – Présence au TD
- ☐ Réussite à l'examen / au partiel
- ☐ Poursuite des études en physique
- ☐ Implication des étudiant.e.s dans la résolution des exercices en TD
- ☐ Sollicitation des enseignant.e.s par les étudiant.e.s (par mail, en TD, dans son bureau, etc.)
- ☐ Autre, précisez

La modalité « TD » vous semble-t-elle remplaçable par
- ☐ Des corrections d'exercices filmées et disponibles *via* l'intranet
- ☐ Des exercices corrigés mis à disposition des étudiants sur l'intranet
- ☐ Du tutorat
- ☐ Les TD ne sont remplaçables par rien d'autre.
- ☐ Autre, précisez

Approuveriez-vous une organisation pédagogique qui repose exclusivement sur des cours-TD (plage de travail au cours de laquelle des exercices alternent avec des points de cours) ?
- ☐ Oui
- ☐ Non
- ☐ Commentaire libre

Approuveriez-vous le fait d'assurer une permanence d'accueil de vos étudiant.e.s (réuni.e.s en groupe restreint de 5 à 10 étudiants) pour répondre à leurs questions, approfondir un point du cours, orienter leurs lectures, etc. ?
- ☐ Oui
- ☐ Non
- ☐ Commentaire libre

### *Les difficultés associées à l'enseignement de la physique*

Que trouvez-vous difficile dans votre métier d'enseignant.e ? (En cas de réponses multiples, numérotez vos choix de 1 à 3).
- ☐ Les carences dans la formation disciplinaire des étudiant.e.s
- ☐ Le nombre d'heures d'enseignement
- ☐ Le manque de formation des enseignant.e.s-chercheur.e.s
- ☐ La charge de travail administratif
- ☐ Le manque d'échanges autour des pratiques d'enseignement
- ☐ Le manque de ressources pédagogiques
- ☐ L'hétérogénéité du niveau des étudiant.e.s
- ☐ Rien n'est difficile
- ☐ Autres, précisez

Selon vous, quelles sont les difficultés principales des étudiant.e.s ? Numérotez les trois principales selon leur ordre d'importance.
- ☐ Le nombre d'heures de présence à l'Université
- ☐ L'insuffisante maîtrise des outils mathématiques
- ☐ L'insuffisance de leurs connaissances en physique
- ☐ Le manque de travail
- ☐ Le manque de curiosité / de motivation
- ☐ Le manque d'autonomie

☐ La difficulté à donner du sens aux savoirs (lois, concepts, principes…) en physique
☐ Autres, précisez

### *L'évaluation des étudiant.e.s*

Parmi les propositions suivantes, quelle est celle qui vous paraît la mieux adaptée à l'évaluation des étudiant.e.s ?
☐ Un partiel et un examen final pour chaque enseignement
☐ Un contrôle continu (avec partiel) et un examen final pour chaque enseignement
☐ Un contrôle continu (sans partiel) et un examen final pour chaque enseignement
☐ Un examen final uniquement
☐ Un contrôle continu uniquement
☐ Aucune de ces propositions ne me paraît adaptée
☐ Autre proposition

Quelles modalités vous semblent adaptées à un contrôle continu efficace ?
☐ Devoirs à la maison
☐ Interrogation écrite en début de TD
☐ Interrogation orale en début de TD
☐ Colles
☐ Le contrôle continu est inutile
☐ Autres, précisez

Certaines universités mettent en place des systèmes de réponses en temps réel nommés « clickers ». Présentés sous forme de petits boitiers individuels, les clickers permettent de connaître de façon quasi-instantanée le profil d'un groupe d'étudiant.e.s soumis à une (ou plusieurs) question(s) à choix multiple. Seriez-vous favorable à la mise en place de cette technologie dans votre enseignement ?
☐ Oui
☐ Non
☐ Commentaire libre

### *L'évaluation des enseignements*

À quelle modalité d'évaluation de votre enseignement seriez-vous favorable ?
☐ Le/la responsable de l'UE distribue et récupère un questionnaire anonyme élaboré par ses soins pour chaque enseignant.e
☐ Le/la responsable de la spécialité distribue et récupère un questionnaire anonyme élaboré par ses soins pour chaque UE
☐ Mon ressenti personnel me suffit
☐ Je distribue moi-même un questionnaire anonyme élaboré par mes soins
☐ Discussion collective avec les étudiants

☐ Discussion individuelle avec un.e étudiant.e
☐ Je ne suis pas favorable à l'évaluation de mon enseignement
☐ Autre, précisez

Faudrait-il procéder à l'évaluation de votre enseignement…
☐ En cours de semestre ?
☐ À la fin du semestre ?
☐ Je n'ai pas d'avis
☐ Commentaire libre

Seriez-vous intéressé.e par une (ou plusieurs) séances d'auto-confrontation (technique d'évaluation audiovisuelle de nature « formative » qui consiste, pour un.e enseignant.e, à se regarder « en train de faire cours » à côté d'un pair venu le filmer)
☐ Oui
☐ Non
☐ Commentaire libre

### *La formation des enseignant.e.s-chercheur.e.s*

Ressentez-vous le besoin d'être formé en tant qu'enseignant.e à l'Université ?
☐ Oui
☐ Non
☐ Commentaire libre

Si des propositions de formations vous étaient offertes, quelles sont celles auxquelles vous souhaiteriez assister ?
☐ Apport d'informations sur le profil des étudiant.e.s qui entrent en L de physique (niveau de connaissances, programmes de lycées…)
☐ Formation aux techniques pédagogiques (gestion d'un groupe d'étudiants de L, favoriser les échanges et la prise de parole des étudiant.e.s, s'initier à la pédagogie de projets,…)
☐ Espaces d'échanges de pratiques entre collègues (de la même discipline, de disciplines différentes)
☐ Apport d'informations sur les difficultés de nature cognitive associées à l'enseignement des certains concepts, de certaines lois en physique
☐ Autre, précisez

Quel temps seriez-vous prêt à consacrer à une telle formation ?
☐ 3h/mois
☐ 3 jours/an
☐ 3 jours tous les 3 ans
☐ Autre

### *Commentaires libres généraux*